# Photo-Hall, photorefractive and photomagnetoelectric effects in tungsten bronzes and related tetragonal ferroelectrics

I. Tekaya[1], A. Tekaya[1], B. Maximin[2]

[1] *Laboratoire de Physique de la Matière Condensée, Université de Picardie Jules Verne, 33 rue Saint-Leu, 80039 Amiens Cedex, France*
[2] *Physique des Systèmes Complexes, Université de Picardie Jules Verne, 33 rue Saint-Leu, 80039 Amiens Cedex, France*

We present an extensive study of the electric, magnetic and elastic responses of tetragonal ferroelectrics under illumination, using a new theory intertwining material and wave symmetries. Optical rectification, photomagnetic, photovoltaic and phototoroidal vector responses are worked out as functions of the wave vector and wave polarization directions. Second-order response tensors associated with photoelastic, photomagnetoelectric, photorefractive effects and photoconductivity are described. We discuss in detail the photo-Hall and a new non-linear optical effect in tungsten bronzes. Finally, we compare the properties of tetragonal materials with those of hexagonal, orthorhombic and trigonal ferroelectrics previously reported in the literature.

## I. INTRODUCTION

The family of tetragonal tungsten bronze (TTB) materials, evidenced since 1949 and intensively studied both as single crystals and as ceramics [1—7], exhibits exceptional ferroelectric behaviors [8,9], together with non-linear optical properties. These features make them interesting materials for electro-optics and optoelectronics [1,10—16] as well as promising candidates for computer memory applications based on resistive switching phenomena [17,18]. An additional advantage of these materials is the large variety of atoms present in their cell, which permit many chemical substitutions giving rise, for instance, to the possibility to give magnetic properties to these structures. Thus, bronze tungstens potentially become multiferroic if they can undergo a magnetic ordering transition and can acquire magnetic properties even in their non-magnetic phases. Moreover, their stability under intense laser light and their high non-linear coefficients open the possibility both for using illumination to efficiently probe these materials, but also for exploiting them in order to evidence new phenomena resulting from their coupling with high-intensity light beams.

We investigate theoretically such phenomena on applying an approach recently proposed [19] to predict the tensorial response of single crystals to linearly polarized light waves. We will focus attention only on the physical effects involving four types of vectors (section 3) and the four types of second-rank tensors (section 4), symmetric or antisymmetric with respect to time and space reversals.

We illustrate our results with the help of a limited number of examples: optical rectification [20], photomagnetic [21], phototoroidal [22,23] and photovoltaic effects for vector responses, on the one hand, and photoelastic [24], photorefractive [25—27], photomagnetoelectric, photoconductivity [28] and photo-Hall effects [29,30] for tensors, on the other hand. The former permit to calculate the polarization, magnetization, toroidal moment and electric currents induced (or modified for the polarization in ferroelectric phases) under illumination. The latter predict the light-induced elastic deformation and the modifications of various response coefficients: optic tensor, magneto-electric tensor, conductivity and a Hall-type tensor. We pay special attention to the distinction between the effects due to dissipative and non-dissipative processes [19,31], respectively, which exhibit distinct symmetry properties. This theoretical approach has a phenomenological character and it is only based on the symmetry of the light beam and that of the crystal. Thus, our results have a wide range of applications, since they apply to any material with tetragonal (more precisely with point group $C_{4v}$ or $D_{4h}$) and possibly orthorhombic ($C_{2v}$) symmetries in their phase diagrams.

We use two TTBs, namely the lead potassium niobate $Pb_2KNb_5O_{15}$ (PKN) and the gadolinium potassium niobate $GdK_2Nb_5O_{15}$ (GKN), as typical examples of such materials. PKN is orthorhombic ferroelectric at room temperature [3]. Its space group is Cm2m and it undergoes a phase transition at $T_C = 723\ K$, becoming paraelectric with P4/mbm symmetry. Its high Curie temperature proves very useful for optical applications such as waveguides and birefringence [32]. GKN, also ferroelectric at room temperature with space group P4bm [5], undergoes two phase transitions at 511 K and 648 K related to the ferroelectric-antiferroelectric-paraelectric phases [33]. Indeed, its tetragonal space group in the ferroelectric phase is P4bm and becomes P4nc in the antiferroelectric phase, then switching to P4/mbm in the paraelectric phase [5]. The full polymorphism of PKN and GKN can be summarized with only three point groups: the tetragonal 4/mmm ($D_{4h}$), 4mm ($C_{4v}$) groups and the orthorhombic 2mm ($C_{2v}$) group. The global structure of TTBs is presented on figure 1.

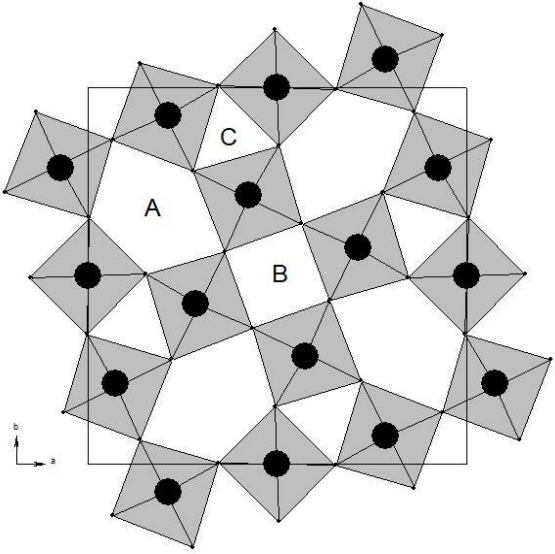

**FIG. 1**. Schematic projection of the structure of tetragonal tungsten bronzes on the $a-b$ plane, showing the pentagonal sites (A), the square sites (B) and the triangular sites (C).

We expose the theoretical formalism in section 2. Then we apply this formalism to vectors and second-order tensors in sections 3 and 4. Finally, we discuss our results and compare them with other materials in section 5.

## II. WIGNER EXPANSION

The theoretical approach we use for studying illumination effects has been recently developed [19] in order to determine the light-induced modifications of physical tensors as functions of the wave vector and wave polarization orientations. Its formalism, based on expansions of the response coefficients in Wigner functions [34], was previously developed [35] within the framework of nematic liquid crystals, and was then applied to multiferroic materials such as the ferroelectric trigonal $LiNbO_3$ [19], the photovoltaic orthorhombic $KBiFe_2O_5$ [36] and the rhombohedral monoclinic $BiFeO_3$ [37]. We extend for the first time this analysis to the class of tetragonal materials. The absence of magnetic structures in their phase diagrams allowed us to slightly simplify the formalism of [19].

In a single crystal of the $D_{4h}$ phase, it is sufficient to consider a single domain. We orient this domain in the laboratory frame $(x, y, z)$ in such a way that the fourfold rotation axis be parallel to $z$ and one of the $\sigma_v$-type mirror planes be normal to $x$. In the tetragonal ferroelectric phase ($C_{4v}$), one has to consider simultaneously two domains, since they respond differently to the light beam. In one domain (domain 1) we proceed as in the $D_{4h}$ phase. The other domain (domain 2) is defined by the application of space inversion $I$ to the first domain. We associate an integer number $\kappa_I$ to each domain, where $\kappa_I = +1$ for the domain 1, and $\kappa_I = -1$ to the domain 2. Of course, the spontaneous polarization is reversed between the two domains.

We then apply the light beam to each domain and probe the modifications of any tensor $\Sigma$ induced by the light beam. The stationary values of this tensor depend then on $\kappa_I$, on the one hand, and to the orientation of the light beam on the other hand. This orientation is characterized by the three Euler angles $\alpha, \beta, \gamma$ that permit to rotate the laboratory frame $x, y, z$ into the (normalized) wave frame $e, b, k$, where $e$ and $b$ are unit vectors proportional to the electric and magnetic polarizations of the linearly polarized wave, and $k$ is proportional to its wave vector. The cartesian components of the response tensor are thus functions $\Sigma^A(\kappa_I, \alpha, \beta, \gamma)$ of the previous parameters. They can then be expanded into Wigner spherical functions $D_{mp}^L(\alpha, \beta, \gamma)$:

$$\Sigma^A(\kappa, \alpha, \beta, \gamma) =$$
$$\sum_{L=0}^{+\infty} \sum_{m,p=L}^{L} \{{}^A K_L^{mp}(0) + \kappa_I {}^A K_L^{mp}(2)\} D_{mp}^L(\alpha, \beta, \gamma) \quad (1)$$

where the ${}^A K_L^{mp}(s = 0,2)$ are complex phenomenological coefficients. $A$ is a tensorial multi-index ($A = i = x, y, z$ for vector responses and $A = (i,j)$ for second-order tensors), $L = 0, 1, 2, \ldots$ and $-L \leq m, p \leq L$ are integer indices similar to the $L$ and $m$ indices of the spherical harmonics $Y_L^m$. Therefore, expansion (1) appears to be analogous to a usual expansion of a function depending on the two spherical angles $\theta, \varphi$ into $Y_L^m$. Accordingly, increasing $L$ amounts to improve the angular precision of Eq. 1. We will use in the sequel a simple approximation that consists in restricting expansion (1) to $L = 0$ and $1$, which provides the main angular variations of $K$. The same restriction was used in [36] for $KBiFe_2O_5$ and in [37] for $BiFeO_3$, whereas a more accurate second-order approximation ($L = 0,1,2$) was used in [19] for $LiNbO_3$.

Even with small values of $L$ the number of independent coefficients ${}^A K_L^{mp}(s)$ is large. It can be drastically reduced when considering the symmetry group of the wave together with that of the crystal. Each generator of these two groups acts linearly on the coefficients ${}^A K_L^{mp}(s)$ and thus provides a linear equation for these parameters (the so-called external (wave) and internal (crystal) selection rules of [19]). The external rules read:

$$\begin{cases} {}^A K_L^{np\,*} = (-1)^{n-p\,A} K_L^{-n\,-p} \\ {}^A K_L^{n\,2p+1} = 0 \\ (-1)^{L+p\,A} K_L^{n\,-p} \varphi_s = \tau_I {}^A K_L^{np} \\ (-1)^{L+s\,A} K_L^{n\,-p} = \tau_T {}^A K_L^{np} \end{cases} \quad (2)$$

where $\varphi_0 = 1$ and $\varphi_2 = -1$, $\tau_I = +1$ or $-1$ respectively for a symmetric or antisymmetric tensor under space inversion, and $\tau_T = +1$ or $-1$ in the same manner for time reversal. The fourth equation is related to time reversal symmetry and permits to distinguish dissipative from non-dissipative terms in Eq. 1 [19,20].

The internal rule reads, for rotation symmetries:

$$R_B^A \sum_{m=-L}^{+L} {}^B K_L^{mn}(s) D_{mp}^L(R^{-1}) = {}^A K_L^{np}(s) \quad (3)$$

where $R_B^A = R_j^i$ is the rotation matrix in cartesian coordinates (for vectors) or $R_j^i R_l^k$ (for tensors), whereas space inversion transforms $\kappa_I$ into $-\kappa_I$ and time reversal transforms $\kappa_T$ into $-\kappa_T$.

Solving the corresponding set of internal and external equations allows to determine the general form of the tensor $\Sigma$ as a linear combination of a reduced number of real parameters (Tables I—VIII). This form depends on the order and the type of the tensor and on the crystal point group.

## III. VECTORIAL RESPONSES

Illumination can quantitatively and qualitatively modify any physical quantity that is characteristic of the equilibrium state of the material. For instance, in an illuminated stationary state, some vector quantities may appear: current density $j$ (photovoltaic effect), electric polarization $P$ (optical rectification) and magnetization $M$ (photomagnetic effect). $j$ and $M$ vanish in all the phases when the light is turned off, while $P$ preexists in the ferroelectric phases $C_{4v}$ and $C_{2v}$. These three vectors exhibit distinct behaviors when they are illuminated because they obey distinct transformation laws with respect to space $(I)$ and time $(R)$ reversals. $j$ is anti-symmetric under both $I$ and $R$ (type [-1,-1], $P$ is symmetric for $R$ and antisymmetric for $I$ (type [-1,1]), and conversely for $M$ (type [1,-1]). For completeness, we will also describe the behavior of a vector $A$, symmetric with respect to $I$ and $R$ (type [1,1]), which has however less physical examples than $j$, $M$ and $P$. Let us notice that in a magnetic insulator, $j$ can be replaced with the toroidal moment $T$, which describes a combination of spins arising spontaneously in specific antiferromagnets and has exactly the same symmetry properties as $j$. Accordingly, it formally exhibits the same physical behavior, so that all the qualitative conclusions we will get for $j$ in a conductor or a semi-conductor can be immediately translated for $T$ in the non-ordered phases of a magnetic insulator.

We can calculate the values of these four vectors as presented in section 2 (Eq. 1,2,3). The results are shown in Tab. 1—4 for the three phases of tungsten bronzes, where Wigner expansions are restricted to the first order in $L (L \leq 1)$.

**TABLE I.** Type [1,1] vector (symmetric under both space and time reversals) for crystals with point groups $D_{4h}$, $C_{4v}$ and $C_{2v}$. The components are expressed as functions of the spherical angles $\alpha, \beta$ of the wave vector with respect to the crystal axes. The phenomenological coefficients $a, A, B \ldots$ are arbitrary real quantities. Small letters $(c)$ are used to present the non-dissipative terms whereas capital letters $(A, B)$ present the dissipative ones. Column 1: Point group of the phase. Column 2: Equilibrium value of $A$ without illumination. Column 3: Additional stationary value of $A$ under illumination.

| Crystal point group | Spontaneous $A$ [1,1] | Illuminated $A$ [1,1] |
|---|---|---|
| $D_{4h}$ | $\begin{pmatrix} 0 \\ 0 \\ 0 \end{pmatrix}$ | $\begin{pmatrix} 0 \\ 0 \\ c \end{pmatrix}$ |
| $C_{4v}$ | $\begin{pmatrix} 0 \\ 0 \\ A_0 \end{pmatrix}$ | $\begin{pmatrix} \kappa_I A \sin\beta \sin\alpha \\ \kappa_I A \sin\beta \cos\alpha \\ c \end{pmatrix}$ |
| $C_{2v}$ | $\begin{pmatrix} 0 \\ 0 \\ A_0 \end{pmatrix}$ | $\begin{pmatrix} \kappa_I A \sin\beta \sin\alpha \\ \kappa_I B \sin\beta \cos\alpha \\ c \end{pmatrix}$ |

**TABLE II.** Polarization $P$ (type [-1,1]), for crystals with point groups $D_{4h}$, $C_{4v}$ and $C_{2v}$. The components are expressed as functions of the spherical angles α, β of the wave vector with respect to the crystal axes. The phenomenological coefficients $a, A, B \ldots$ are arbitrary real quantities. Small letters $(c)$ are used to present the non-dissipative terms whereas capital letters $(A, B)$ present the dissipative ones. Column 1: Point group of the phase. Column 2: Equilibrium value of $P$ without illumination. Column 3: Additional stationary value of $P$ under illumination.

| Crystal point group | Spontaneous $P$ [-1,1] | Illuminated $P$ [-1,1] |
|---|---|---|
| $D_{4h}$ | $\begin{pmatrix} 0 \\ 0 \\ 0 \end{pmatrix}$ | $\begin{pmatrix} B \sin\beta \cos\alpha \\ -B \sin\beta \sin\alpha \\ 0 \end{pmatrix}$ |
| $C_{4v}$ | $\begin{pmatrix} 0 \\ 0 \\ A_0 \end{pmatrix}$ | $\begin{pmatrix} A \sin\beta \sin\alpha \\ -A \sin\beta \cos\alpha \\ C \cos\beta \end{pmatrix}$ |
| $C_{2v}$ | $\begin{pmatrix} 0 \\ 0 \\ A_0 \end{pmatrix}$ | $\begin{pmatrix} A \sin\beta \sin\alpha \\ -B \sin\beta \cos\alpha \\ C \cos\beta \end{pmatrix}$ |

### A. Optical rectification

Table II shows that in the three phases of the tungsten bronzes, the additional polarization induced by light corresponds only to dissipative terms, so that they are negligible at low intensity. In the paraelectric $(D_{4h})$ phase, the induced polarization is always in the $x - y$ plane (Fig. 2). The $x - y$ plane projection of $P$ is parallel to the wave vector, except in the orthorhombic phase (Fig. 4). When the wave vector is parallel to $z$, the in-plane component of $P$ vanishes, while it is maximum along $z$ in the ferroelectric phases ($C_{4v}$ and $C_{2v}$). When the wave vector is normal to $z$, the polarization also lies within the $x - y$ plane (Fig. 2—4).

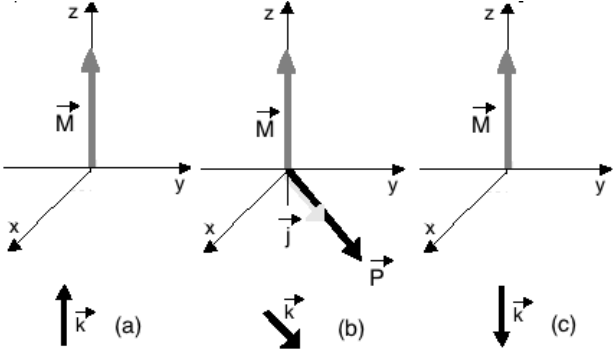

**FIG. 2.** Induced polarization **P**, magnetization **M** and electric current density **j** in the paraelectric phase $D_{4h}$. (a) Wave vector **k** parallel to $+z$. (b) **k** in the $x-y$ plane. (c) **k** parallel to $-z$.

**TABLE III.** Magnetization $M$ (type [1,-1]), for crystals with point groups $D_{4h}$, $C_{4v}$ and $C_{2v}$. The components are expressed as functions of the spherical angles $\alpha, \beta$ of the wave vector with respect to the crystal axes. The phenomenological coefficients $a, A, B...$ are arbitrary real quantities. Small letters ($c$) are used to present the non-dissipative terms whereas capital letters ($A, B$) present the dissipative ones. Column 1: Point group of the phase. Column 2: Equilibrium value of $M$ without illumination. Column 3: Additional stationary value of $M$ under illumination.

| Crystal point group | Spontaneous $M$ [1,1] | Illuminated $M$ [1,1] |
|---|---|---|
| $D_{4h}$ | $\begin{pmatrix} 0 \\ 0 \\ 0 \end{pmatrix}$ | $\begin{pmatrix} 0 \\ 0 \\ C \end{pmatrix}$ |
| $C_{4v}$ | $\begin{pmatrix} 0 \\ 0 \\ 0 \end{pmatrix}$ | $\begin{pmatrix} \kappa_I a \sin\beta \sin\alpha \\ \kappa_I a \sin\beta \cos\alpha \\ C \end{pmatrix}$ |
| $C_{2v}$ | $\begin{pmatrix} 0 \\ 0 \\ 0 \end{pmatrix}$ | $\begin{pmatrix} \kappa_I a \sin\beta \sin\alpha \\ \kappa_I b \sin\beta \cos\alpha \\ C \end{pmatrix}$ |

### B. Photomagnetic effect

The spontaneous magnetization vanishes in all the considered phases. The non-dissipative terms that appear under illumination in the ferroelectric phases are normal to the wave vector. They vanish when **k** is parallel to $z$, and are maximum when **k** is normal to $z$. Their direction is reversed when one considers two domains with opposite spontaneous polarizations (Fig. 2—4).

The $z$-component of **M** is always dissipative and isotropic: it does not depend on the propagation direction of the wave. Moreover, it is independent on the sense of the spontaneous polarization (Fig. 3).

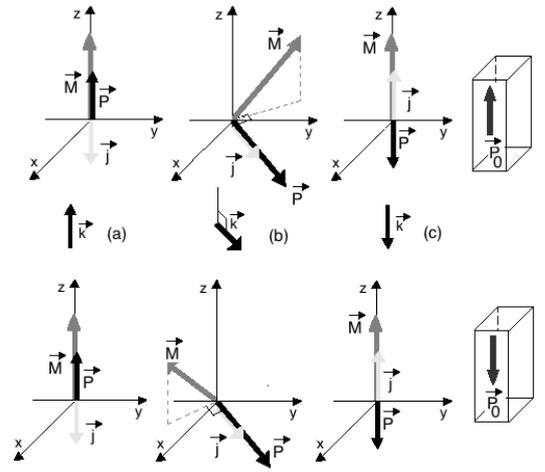

**FIG. 3.** Optical rectification, photomagnetic and photovoltaic effects in the tetragonal ferroelectric ($C_{4v}$) phase under illumination. **k** is the wave vector, **P** the induced polarization, **M** the induced magnetization and **j** the induced electric current. Upper row: in the tetragonal domain with up spontaneous polarization $P_0$. Down row: in the tetragonal domain with down spontaneous polarization $P_0$. (a) **k** parallel to $+z$, (b) **k** normal to $z$. (c) **k** parallel to $-z$.

**TABLE IV.** Electric current $j$ and toroidal moment $T$ (type [-1,-1]), for crystals with point groups $D_{4h}$, $C_{4v}$ and $C_{2v}$. The components are expressed as functions of the spherical angles $\alpha, \beta$ of the wave vector with respect to the crystal axes. The phenomenological coefficients $a, A, B...$ are arbitrary real quantities. Small letters ($c$) are used to present the non-dissipative terms whereas capital letters ($A, B$) present the dissipative ones. Column 1: Point group of the phase. Column 2: Equilibrium value of $j, T$ without illumination. Column 3: Additional stationary value of $j, T$ under illumination.

| Crystal point group | Spontaneous $j, T$ [-1,-1] | Illuminated $j, T$ [-1,-1] |
|---|---|---|
| $D_{4h}$ | $\begin{pmatrix} 0 \\ 0 \\ 0 \end{pmatrix}$ | $\begin{pmatrix} b\sin\beta\cos\alpha \\ -b\sin\beta\sin\alpha \\ 0 \end{pmatrix}$ |
| $C_{4v}$ | $\begin{pmatrix} 0 \\ 0 \\ 0 \end{pmatrix}$ | $\begin{pmatrix} a\sin\beta\sin\alpha \\ -a\sin\beta\cos\alpha \\ c\cos\beta \end{pmatrix}$ |
| $C_{2v}$ | $\begin{pmatrix} 0 \\ 0 \\ 0 \end{pmatrix}$ | $\begin{pmatrix} a\sin\beta\sin\alpha \\ -b\sin\beta\cos\alpha \\ c\cos\beta \end{pmatrix}$ |

### C. Photovoltaic effect

The light-induced current exhibits only non-dissipative terms independent of the ferroelectric domains. The $x-y$ components are normal to the wave vector, except in the orthorhombic phase. They vanish when $k$ is parallel to $z$ (Fig. 2—4).

The $z$-component of $j$ vanishes in the non-polar phase and when the wave is normal to $z$ in the polar phases. It is maximum when the wave is parallel to $z$. In this case it changes its sign when the sense of propagation is reversed (Fig. 2—4).

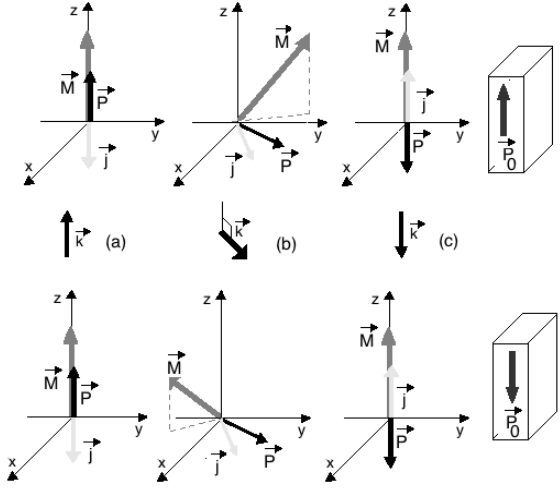

**FIG. 4.** Optical rectification, photomagnetic and photovoltaic effects in the orthorhombic ferroelectric ($C_{2v}$) phase under illumination. $k$ is the wave vector, $P$ the induced polarization, $M$ the induced magnetization and $j$ the induced electric current. Upper row: In one orthorhombic domain with up spontaneous polarization $P_0$. Down row: In the corresponding orthorhombic domain with down spontaneous polarization $P_0$. (a) $k$ parallel to $+z$, (b) $k$ normal to $z$. (c) $k$ parallel to $-z$.

## IV. TENSORIAL RESPONSES

Second-order tensors $^{ij}K$ ($i,j = x,y,z$), which play a central role in material physics as response coefficients associated with various external excitations, may also be classified according to their parity with respect to space and time reversals yielding four types of tensors. Moreover, symmetric ($^{ij}S = {}^{ji}S$) from skew-symmetric ($^{ij}A = -{}^{ji}A$) tensors must be distinguished. Thus, one must evaluate the Wigner forms (Eq. 1) corresponding to eight types of such tensors. For instance, the dielectric constant $^{ij}\varepsilon$ is a (1,1) type symmetric tensor invariant under the two parities, whereas the magnetoelectric response ($^{ij}\alpha$) is a (-1,-1) type tensor antisymmetric under time and space reversals containing both symmetric and skew-symmetric contributions.

Solving the corresponding selections rules provides the Wigner expansions presented in Tables V—VIII up to first order in $L$, where dissipative and non-dissipative terms are distinguished.

**TABLE V.** Matrices of the tensors of type [1,1] (symmetric under both space and time reversals) for crystals with point groups $D_{4h}$, $C_{4v}$ and $C_{2v}$. The response coefficients are expressed as functions of the spherical angles $\alpha, \beta$ describing the orientation of the wave vector with respect to the crystal axes. The phenomenological coefficients $a, f, D, ...$ are arbitrary real quantities. Small letters ($a, f, ...$) are used to present the non-dissipative terms whereas capital letters ($D, \Delta, ...$) present the dissipative ones. Roman and Greek letters represent the symmetric and skew-symmetric parts of the tensors respectively.

| Crystal point group | Response tensors [1,1] |
|---|---|
| $D_{4h}$ | $\begin{pmatrix} a & & \\ & a & \\ & & f \end{pmatrix}$ |
| $C_{4v}$ | $\begin{pmatrix} a & 0 & (D+\Delta)\kappa_I \sin\beta \cos\alpha \\ 0 & a & -(D+\Delta)\kappa_I \sin\beta \sin\alpha \\ (D-\Delta)\kappa_I \sin\beta \cos\alpha & (-D+\Delta)\kappa_I \sin\beta \sin\alpha & f \end{pmatrix}$ |
| $C_{2v}$ | $\begin{pmatrix} a + A\cos\beta\,\kappa_I & 0 & (D+\Delta)\kappa_I \sin\beta \cos\alpha \\ 0 & c + C\cos\beta\,\kappa_I & (E+\mathrm{E})\kappa_I \sin\beta \sin\alpha \\ (D-\Delta)\kappa_I \sin\beta \cos\alpha & (E+\mathrm{E})\kappa_I \sin\beta \sin\alpha & f + F\cos\beta\,\kappa_I \end{pmatrix}$ |

**TABLE VI.** Matrices of the tensors of type [-1,1] (antisymmetric under space reversal and symmetric under time-reversals) for crystals with point groups $D_{4h}$, $C_{4v}$ and $C_{2v}$. The response coefficients are expressed as functions of the spherical angles $\alpha, \beta$ describing the orientation of the wave vector with respect to the crystal axes. The phenomenological coefficients $a, f, D, ...$ are arbitrary real quantities. Small letters ($a, f, ...$) are used to present the non-dissipative terms whereas capital letters ($D, \Delta, ...$) present the dissipative ones. Roman and Greek letters represent the symmetric and skew-symmetric parts of the tensors respectively.

| Crystal point group | Response tensors [-1,1] |
|---|---|
| $D_{4h}$ | $\begin{pmatrix} 0 & \Gamma\cos\beta & (D+\Delta)\sin\beta\sin\alpha \\ -\Gamma\cos\beta & 0 & -(D+\Delta)\sin\beta\cos\alpha \\ (D-\Delta)\sin\beta\sin\alpha & (D-\Delta)\sin\beta\cos\alpha & 0 \end{pmatrix}$ |
| $C_{4v}$ | $\begin{pmatrix} 0 & \varsigma\kappa_I + \Gamma\cos\beta & (D+\Delta)\sin\beta\sin\alpha \\ -\varsigma\kappa_I - \Gamma\cos\beta & 0 & -(D+\Delta)\sin\beta\cos\alpha \\ (D-\Delta)\sin\beta\sin\alpha & (-D+\Delta)\sin\beta\cos\alpha & 0 \end{pmatrix}$ |
| $C_{2v}$ | $\begin{pmatrix} 0 & g\kappa_I + (B+\Gamma)\cos\beta & (D+\Delta)\sin\beta\sin\alpha \\ -g\kappa_I + (B-\Gamma)\cos\beta & 0 & (-F+\mathrm{E})\sin\beta\cos\alpha \\ (D-\Delta)\sin\beta\sin\alpha & -(F+\mathrm{E})\sin\beta\cos\alpha & 0 \end{pmatrix}$ |

**TABLE VII.** Matrices of tensors of type [1,-1] (antisymmetric under time reversal and symmetric under space inversion) for crystals with point groups $D_{4h}$, $C_{4v}$ and $C_{2v}$. The response coefficients are expressed as functions of the spherical angles $\alpha, \beta$ describing the orientation of the wave vector with respect to the crystal axes. The phenomenological coefficients $a, f, D, ...$ are arbitrary real quantities. Small letters $(a, f, ...)$ are used to present the non-dissipative terms whereas capital letters $(D, \Delta, ...)$ present the dissipative ones. Roman and Greek letters represent the symmetric and skew-symmetric parts of the tensors respectively.

| Crystal point group | Response tensors [1,-1] |
|---|---|
| $D_{4h}$ | $\begin{pmatrix} A & & \\ & A & \\ & & F \end{pmatrix}$ |
| $C_{4v}$ | $\begin{pmatrix} A & 0 & (d+\delta)\kappa_I \sin\beta \cos\alpha \\ 0 & A & -(d+\delta)\kappa_I \sin\beta \sin\alpha \\ (d-\delta)\kappa_I \sin\beta \cos\alpha & (-d+\delta)\kappa_I \sin\beta \sin\alpha & F \end{pmatrix}$ |
| $C_{2v}$ | $\begin{pmatrix} A + a' \cos\beta \, \kappa_I & 0 & (d+\delta)\kappa_I \sin\beta \cos\alpha \\ 0 & C + c' \cos\beta \, \kappa_I & (e+\varepsilon)\kappa_I \sin\beta \sin\alpha \\ (d-\delta)\kappa_I \sin\beta \cos\alpha & (e-\varepsilon)\kappa_I \sin\beta \sin\alpha & F + f' \cos\beta \, \kappa_I \end{pmatrix}$ |

**TABLE VIII.** Matrices of the tensors of type [-1,-1] (anti-symmetric under both space and time reversals) for crystals with point groups $D_{4h}$, $C_{4v}$ and $C_{2v}$. The response coefficients are expressed as functions of the spherical angles $\alpha, \beta$ describing the orientation of the wave vector with respect to the crystal axes. The phenomenological coefficients $a, f, D, ...$ are arbitrary real quantities. Small letters $(a, f, ...)$ are used to present the non-dissipative terms whereas capital letters $(D, \Delta, ...)$ present the dissipative ones. Roman and Greek letters represent the symmetric and skew-symmetric parts of the tensors respectively.

| Crystal point group | Response tensors [-1,-1] |
|---|---|
| $D_{4h}$ | $\begin{pmatrix} 0 & \gamma \cos\beta & (d+\delta)\sin\beta \sin\alpha \\ -\gamma \cos\beta & 0 & (d+\delta)\sin\beta \cos\alpha \\ (d-\delta)\sin\beta \cos\alpha & (d-\delta)\sin\beta \sin\alpha & 0 \end{pmatrix}$ |
| $C_{4v}$ | $\begin{pmatrix} A\kappa_I & \gamma \cos\beta & (d+\delta)\sin\beta \sin\alpha \\ -\gamma \cos\beta & A\kappa_I & (d+\delta)\sin\beta \cos\alpha \\ (d-\delta)\sin\beta \cos\alpha & (d-\delta)\sin\beta \sin\alpha & F\kappa_I \end{pmatrix}$ |
| $C_{2v}$ | $\begin{pmatrix} A\kappa_I & \gamma \cos\beta & (d+\delta)\sin\beta \sin\alpha \\ -\gamma \cos\beta & D\kappa_I & (e+\varepsilon)\sin\beta \cos\alpha \\ (d-\delta)\sin\beta \cos\alpha & (e-\varepsilon)\sin\beta \sin\alpha & F\kappa_I \end{pmatrix}$ |

The phenomenological coefficients $a, b, D, \Delta, ...$ appearing in these tensor matrices are let undetermined by our symmetry analysis. They do not depend on the wave orientation but depend on the beam intensity and frequency and on the material properties and temperature. These tensors allow to foresee a large variety of phenomena that we will illustrate on considering only a few cases in the limit of low beam intensity, when one may restrict the Wigner expansions to their dominant non-dissipative terms. One physical example for each type of tensor will be discussed: strain tensor (type [1,1]), magnetoelectric tensor (type [-1,-1]), electric (type [1,-1]) and magnetic (type [-1,1]) conductivities.

### A. Photoelastic and photorefractive effects

The symmetry breakdown induced by illumination yields the onset of non-trivial components of the strain tensor $^{ij}T$. Its non-dissipative terms can be deduced from Table V:

$$T = \begin{pmatrix} a & & \\ & a & \\ & & f \end{pmatrix} + \begin{pmatrix} m & & \\ & -m & \\ & & 0 \end{pmatrix}$$

where the first matrix corresponds to the induced deformation in the $D_{4h}$ phase whereas the second matrix appears in the $C_{2v}$ phase. The effect is isotropic in the sense that no dependence upon the wave direction is observed. Anisotropic effects appear nevertheless at high intensity (Table V) where we can no longer neglect dissipative terms.

Thus, one sees that in the three phases the illumination effect is trivial, i.e. it does not break the symmetry (at this order of approximation) and cannot be distinguished from thermal expansion. In addition, the effect is exactly the same in the two types of ferroelectric domains in $C_{2v}$ and $C_{4v}$ phases. These conclusions also hold for the dielectric and diamagnetic susceptibility tensors because they are of the same type as the strain tensor. However, one sees in Table VIII that non-dissipative terms have more complex effects in the two ferroelectric phases where shear strain is produced by the light wave, and different effects occur in distinct domains.

The refractive index $n$ is also a type [1,1] second-order tensor and follows the same behavior under illumination as the elastic deformation. Its light-induced modifications define the photorefractive effect, which provides a variety of non-linear optic phenomena. Thus, at the symmetry point of view, the refractive effect is qualitatively trivial when one considers only non-dissipative terms. At larger beam intensity dissipative terms become relevant and can provoke qualitatively non-trivial effects.

The type [1,1] tensor symmetric contribution presented in Table V gives the form of the refractive tensor in the $C_{4v}$ phase that reads, taking into account zero-beam and dissipative digonal terms $(n_x, n_y, n_z)$, on the one hand, and non-dissipative light-induced non-diagonal components, on the other hand:

$$n = \begin{pmatrix} n_x & 0 & D\kappa_I \sin\beta \cos\alpha \\ 0 & n_y & D\kappa_I \sin\beta \sin\alpha \\ D\kappa_I \sin\beta \cos\alpha & D\kappa_I \sin\beta \sin\alpha & n_z \end{pmatrix}$$

Since the effects of the diagonal terms have already been discussed above, let us focus on the symmetry-breaking non-diagonal terms. When the beam is not parallel to $z$ they induce biaxiality together with a rotation of the principal optic axes. In the direction $\mathbf{k'}$ normal to the projection of $\mathbf{k}$ onto the optically isotropic $x-y$ plane (Fig. 5a), the first principal refractive index is not modified by the illumination. The two other indices become (to the second order in the small parameter $D$):

$$n'_x = n_x + \frac{D^2}{n_x - n_z}, n'_z = n_z - \frac{D^2}{n_x - n_z} \quad (4)$$

(assuming $n_z < n_x$), which take the same values in the two domains of polarization ($\kappa_I = \pm 1$). On the other hand, the rotation of the two corresponding principal optic axes around $\mathbf{k'}$ are opposite in these two domains (Fig. 5b). The rotation angle $\alpha$ is given by:

$$\tan \alpha = \frac{D \kappa_I}{n_x - n_z} \quad (5)$$

Thus, even though the light wave propagates along $x$, $y$ or $z$ it is never parallel to any modified principal optic axis ($z'$, $\mathbf{k'_\perp}$ and $\mathbf{k'}$ in Fig. 5a) and splits into an ordinary and an extraordinary wave. Oppositely, an additional test ray with small intensity and traveling parallel to $\mathbf{k'}$ is not splitted.

Similarly, when the second ray is not parallel to $\mathbf{k'}$, it feels the induced biaxiality of the optic tensor and splits into distinct rays. As an illustration of this non-linear optic effect, let us consider the main beam parallel to $x$ and the secondary beam along $y$ (Fig. 5b). The light-modified principal optic axes are (i) along $y$, (ii) in the $x-z$ plane, (iii) rotated by an angle $\theta = \frac{D \kappa_I}{n_x - n_z}$ with respect to $x$ and $z$ (assuming $D$ small). Then, no splitting is observed for the secondary wave, whereas when it is along $z$ (Fig. 5c) it splits into an ordinary wave that remains parallel to $z$ and an extraordinary wave that is deviated in one sense when the spontaneous polarization is along $+z$, and to the opposite angle when the spontaneous polarization is along $-z$. Analogously, changing the sense of $\mathbf{k}$ reverses the deviation angle of the extraordinary wave. Since the induced biaxiality is a second-order effect in the small parameter $D$ (Eq. 4) whereas the optic axis rotation is a first-order effect (Eq. 5), one may neglect the former. The refracted extraordinary wave lies then in the $x-z$ plane and its direction with respect to $z$ is inclined by an angle:

$$\phi_{inc} = D \kappa_I \quad (6)$$

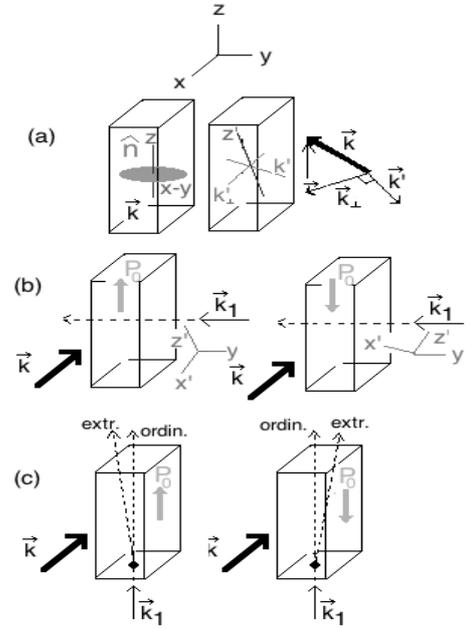

**FIG. 5.** (a) Optic tensor $n$ without (left) and with (right) illumination. $\mathbf{k}$ is the wave vector of the main beam. $\mathbf{k_\perp}$ is its projection onto the $x-y$ plane, $\mathbf{k'}$ the vector of the $x-y$ plane normal to $\mathbf{k_\perp}$. The principal axes of $n$ under illumination are $\mathbf{k'}$ and two vectors $z'$ and $\mathbf{k'_\perp}$ slightly rotated with respect to $z$ and $\mathbf{k_\perp}$. (b) $\mathbf{k}$ is parallel to $\mathbf{x}$ and the wave vector $\mathbf{k_1}$ of the test beam parallel to $\mathbf{y}$. The principal axes of $n$ are $\mathbf{y}$ and $x', z'$ rotated in the $x-z$ plane clockwise if the spontaneous polarization $\mathbf{P_0}$ is up, and anticlockwise if $\mathbf{P_0}$ is down. The test beam is not refracted. (c) $\mathbf{k}$ is parallel to $x$ and the wave vector $\mathbf{k_1}$ of the test beam parallel to $z$. The test beam is refracted into an ordinary wave along $z$ and an extraordinary lying in the $x-z$ plane wave deviated with respect to $z$ by an angle depending on the sense of $\mathbf{P_0}$.

### B. Photomagnetoelectric effect

Under an applied magnetic field $\mathbf{B}$ an additional macroscopic polarization $\Delta \mathbf{P}$ may be induced proportionally to the field (and reciprocally a magnetization may appear under an applied electric field): $\Delta \mathbf{P} = \chi \mathbf{B}$. The corresponding magnetoelectric tensor $\chi$ is of type [-1,-1]. In non-magnetically ordered phases the presence of time reversal in the macroscopic symmetry group of the crystal prevents the apparition of the magnetoelectric effect: $\chi = 0$. Nevertheless, even in these materials the time reversal symmetry is broken by illumination, yielding non-zero magnetoelectric response, given in the studied phases by (see Table VIII):

$$\begin{pmatrix} 0 & \gamma \cos \beta & (d+\delta) \sin \beta \sin \alpha \\ -\gamma \cos \beta & 0 & (d+\delta) \sin \beta \cos \alpha \\ (d-\delta) \sin \beta \cos \alpha & (d-\delta) \sin \beta \sin \alpha & 0 \end{pmatrix}$$
$$+ \begin{pmatrix} 0 & 0 & (m+\mu) \sin \beta \sin \alpha \\ 0 & 0 & -(m+\mu) \sin \beta \cos \alpha \\ (m-\mu) \sin \beta \cos \alpha & (-m+\mu) \sin \beta \sin \alpha & 0 \end{pmatrix}$$

where the first matrix represents the tensor in the $D_{4h}$ and $C_{4v}$ phases, whereas the second matrix provides the components onsetting in the $C_{2v}$ phase. Thus, a magnetoelectric effect is present in all the phases under illumination, even at low intensity.

Only dissipative terms, distinguishing the two types of ferroelectric domains, allow to distinguish the $D_{4h}$ from the $C_{4v}$ phase. One notices that in the three phases the illumination has important symmetry breaking effects. The absence of diagonal components shows that applying the magnetic field along a given direction yields no additional polarization along the same direction (however, dissipative terms can give rise to such a result).

When the wave travels in parallel to the fourfold symmetry axis of the crystal ($z$-direction), the magnetoelectric tensor becomes:

$$\begin{pmatrix} 0 & \gamma & 0 \\ -\gamma & 0 & 0 \\ 0 & 0 & 0 \end{pmatrix}$$

Applying the magnetic field $B$ along $+x$ yields the polarization $\Delta P$ along $-y$, while applying $B$ along $+y$ yields $\Delta P$ along $+x$, with the same amplitude for $\Delta P$ in both cases. No effect is present when applying $B$ parallel to $z$.

Conversely, when the wave travels perpendicularly to $z$, the tensor becomes:

$$\begin{pmatrix} 0 & 0 & (d+\delta)\sin\alpha \\ 0 & 0 & (d+\delta)\cos\alpha \\ (d-\delta)\cos\alpha & (d-\delta)\sin\alpha & 0 \end{pmatrix}$$
$$+\begin{pmatrix} 0 & 0 & (m+\mu)\sin\alpha \\ 0 & 0 & -(m+\mu)\cos\alpha \\ (m-\mu)\cos\alpha & (-m+\mu)\sin\alpha & 0 \end{pmatrix}$$

Then, applying the field parallel to $z$ yields an additional polarization in the $x-y$ plane. More precisely, in the tetragonal phases (first matrix) the polarization modulus does not depend on the direction of the wave vector $k$ in the $x-y$ plane. Furthermore, its direction is normal to that of the wave (Fig. 6a). Reciprocally, applying the field normal to $z$ yields $\Delta P$ parallel to $z$, with its modulus vanishing if $B$ is normal to $k$ (Fig. 6b).

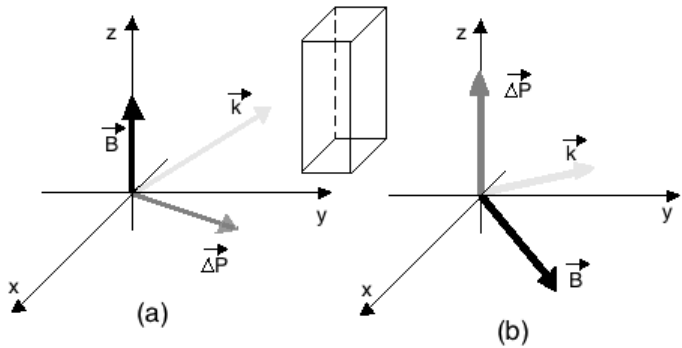

**FIG. 6.** (a) Polarization increment $\Delta P$ induced by a light beam in the $x-y$ crystal plane under applied magnetic field $B$ parallel to $z$ in the tetragonal phases. (b) Polarization increment $\Delta P$ induced by a light beam in the $x-y$ crystal plane under applied magnetic field $B$ normal to $z$ in the tetragonal phases.

The single difference occurring in the orthorhombic phase $C_{2v}$ (second matrix) is that the fourfold symmetry ($x-y$ exchange) is lost. Thus, when the field is parallel to $z$ the additional polarization has a modulus that depends on its direction (Fig. 6), and it is no longer normal to $k$.

Similarly, when $B$ is normal to $z$, $\Delta P$ remains parallel to $z$ but it no longer vanishes when $B$ is normal to $k$.

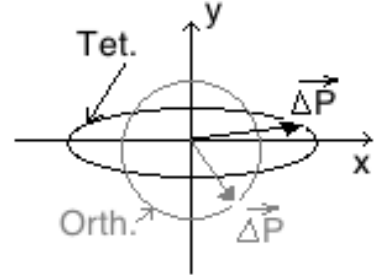

**FIG. 7.** Variations of the polarization increment $\Delta P$ in the $x-y$ plane when the magnetic field is parallel to the $z$ crystal axis and the light wave vector turns in the $x-y$ plane in the tetragonal phases (black ellipse) and in the orthorhombic phase (gray circle).

### C. Photoelectric conductivity

The electric conductivity $\sigma$ is a symmetric tensor of type [1,-1] that is modified by the light beam. Table IX shows the form of $\sigma$ with and without light (neglecting non-dissipative terms in the latter case).

**TABLE IX.** Electric conductivity tensor without (second column) and with (third column) illumination in the three phases. The illumination contributions contain only the non-dissipative terms.

| Crystal point group | Spontaneous $\sigma$ [1,-1] | Illuminated $\sigma$ [1,-1] |
|---|---|---|
| $D_{4h}$ | $\begin{pmatrix} \sigma_x & & \\ & \sigma_y & \\ & & \sigma_z \end{pmatrix}$ | $\begin{pmatrix} 0 & 0 & 0 \\ 0 & 0 & 0 \\ 0 & 0 & 0 \end{pmatrix}$ |
| $C_{4v}$ | $\begin{pmatrix} \sigma_x & & \\ & \sigma_y & \\ & & \sigma_z \end{pmatrix}$ | $d\,\kappa_I \sin\beta \begin{pmatrix} 0 & 0 & \cos\alpha \\ 0 & 0 & -\sin\alpha \\ \cos\alpha & -\sin\alpha & 0 \end{pmatrix}$ |
| $C_{2v}$ | $\begin{pmatrix} \sigma_x & & \\ & \sigma_y & \\ & & \sigma_z \end{pmatrix}$ | $\kappa_I \begin{pmatrix} a'\cos\beta & 0 & d\sin\beta\cos\alpha \\ 0 & c'\cos\beta & e\sin\beta\sin\alpha \\ d\sin\beta\cos\alpha & e\sin\beta\sin\alpha & f'\cos\beta \end{pmatrix}$ |

In the non-polar phase $D_{4h}$ no contribution to the conductivity is produced by the beam within the non-dissipative approximation. In the ferroelectric phase the beam breaks the symmetry of the tensor, which changes of sign in two domains of opposite spontaneous polarizations. Thus, the light-induced current increment $\Delta j$ reverses its sense in the two domains submitted to the same electric field (while the zero-beam contribution is not changed).

In the $C_{4v}$ phase, the tensor vanishes if the beam is parallel to $z$, whereas it is maximal when the beam is in the $x-y$ plane. Moreover, when the beam is not parallel to $z$, the effect does not depend qualitatively on the beam orientation: when the electric field $E$ is in the $x-y$ plane, $\Delta \boldsymbol{j}$ is along $z$ (Fig. 8a), and, reciprocally, when the field is along $z$, $\Delta \boldsymbol{j}$ is in the $x-y$ plane (Fig. 8b). In addition, in the latter case $\Delta \boldsymbol{j}$ is parallel to the in-plane projection of the wave vector $\boldsymbol{k}$, whereas in the former case $\Delta \boldsymbol{j}$ vanishes when $\boldsymbol{E}$ is normal to this projection.

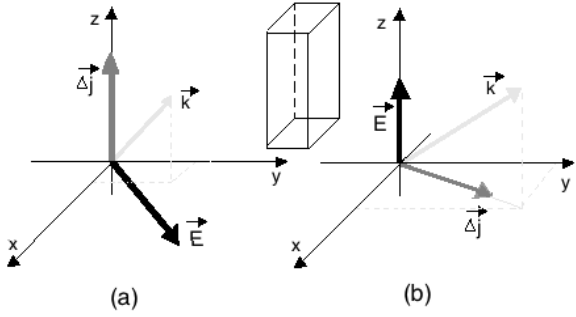

**FIG. 8.** Current increment $\Delta \boldsymbol{j}$ induced by a light beam with wave vector $\boldsymbol{k}$ under applied electric field $\boldsymbol{E}$ in the tetragonal phases. (a) $\boldsymbol{E}$ normal to $z$. (b) $\boldsymbol{E}$ parallel to $z$.

In the orthorhombic phase, the situation is similar to that in $C_{4v}$ when $\boldsymbol{k}$ is normal to $z$. Conversely, when $\boldsymbol{k}$ is parallel to $z$ applying $\boldsymbol{E}$ parallel to $x$, $y$ or $z$ yields $\boldsymbol{j}$ parallel to $\boldsymbol{E}$, as without light (except the reversion of $\Delta \boldsymbol{j}$ in opposite domains). For an arbitrary direction of $\boldsymbol{k}$ the various contributions described here above appear simultaneously.

### D. Photo-Hall effect and magnetoresistance

Magnetic effects are usually observed in metals submitted to a DC electric voltage generating a primary (longitudinal) current, which can either be modified by the applied magnetic field $\boldsymbol{B}$ (magnetoresistance), or give rise to a transverse electric field compensating the transverse current (Hall effect). Such phenomena can also happen in the absence of applied voltage when the material is illuminated. The initial current $\boldsymbol{j}_{ph}$ submitted to the magnetic force, is then induced by the electrons photo-excited in the conduction band and then accelerated by the internal electric field $E_{int}$ of the ferroelectric materials (photovoltaic effect). This process provides an electric current approximately directed along the internal electric field. However, an additional mechanism participates to this effect: the electrons of the material are directly accelerated by photon absorption (photoelectric effect) giving rise to a current roughly directed along the light wavevector. In these two processes, the anisotropy of the material actually deviates the current direction with respect to $E_{int}$ and $\boldsymbol{k}$. All these effects are considered within our formalism, by introducing a new conductivity tensor $\sigma_m$:

$$\boldsymbol{j} = \boldsymbol{j}_{ph} + \sigma_m \boldsymbol{B} \qquad (7)$$

where $\boldsymbol{j}_{ph}$ is the photovoltaic current studied in section 3.3 and the « photomagnetic conductivity » $\sigma_m$ is a symmetric type [-1,1] tensor.

Let us use a simple academic model to illustrate Eq. 7. In an isotropic semiconductor in a stationary state where one neglects the photoelectric part of the induced current, the optical rectification and the photomagnetization (giving rise to higher-order effects), the electron acceleration vanishes so that its velocity obeys

$$\boldsymbol{v} = \frac{e\tau}{m} \boldsymbol{B} \times \boldsymbol{v} + \frac{\boldsymbol{j}_{ph}}{ne}$$

where $\tau$ is the relaxation time [38]. Solving this equation together with Eq. 7 at low magnetic fields yields:

$$\boldsymbol{j}_{ph} = \frac{ne^2\tau}{m} \boldsymbol{E}_{int} \qquad (8)$$

that is, a Drude-type photocurrent ($n$ is the electron density in the conduction band) and yields the following form of the tensor $\sigma_m$:

$$\sigma_m = \frac{ne^3\tau^2}{m^2} \begin{pmatrix} 0 & -E_z^{int} & E_y^{int} \\ E_z^{int} & 0 & -E_x^{int} \\ -E_y^{int} & E_x^{int} & 0 \end{pmatrix} \qquad (9)$$

$\sigma_m$ depends on the symmetry of the material via the internal electric field components, and on the light wave intensity and direction via the light-excited charge carrier density $n$. In a more realistic model, the crystal anisotropy transforms the relaxation time $\tau$ into a symmetric tensor proportional to the zero-beam conductivity. Thus $\sigma_m$ in Eq. 9 becomes the product of a symmetric ($\tau^2$) with an antisymmetric ($\boldsymbol{E}^{int} \times$) tensors, so that it contains both symmetric and antisymmetric contributions. When in addition, one takes into account the photoinduced parts of the internal electric and magnetic fields and the exact form of $\boldsymbol{j}_{ph}$, $\sigma_m$ takes a more intricate form that we will now analyze at the phenomenological level.

For clarity, let us analyze the effects of the symmetric and antisymmetric contributions to $\sigma_m$ separately. Table X shows the forms of the antisymmetric contributions to $\sigma_m$ in the three phases.

**TABLE X.** Antisymmetric contributions to the photomagnetic conductivity tensor in the three phases.

| Crystal point group | $\sigma_m$ |
|---|---|
| $D_{4h}$ | $\begin{pmatrix} 0 & \Gamma \cos\beta & \Delta \sin\beta \sin\alpha \\ -\Gamma \cos\beta & 0 & -\Delta \sin\beta \cos\alpha \\ -\Delta \sin\beta \sin\alpha & \Delta \sin\beta \cos\alpha & 0 \end{pmatrix}$ |
| $C_{4v}$ | $\begin{pmatrix} 0 & \varsigma\kappa_I + \Gamma \cos\beta & \Delta \sin\beta \sin\alpha \\ -\varsigma\kappa_I - \Gamma \cos\beta & 0 & -\Delta \sin\beta \cos\alpha \\ -\Delta \sin\beta \sin\alpha & \Delta \sin\beta \cos\alpha & 0 \end{pmatrix}$ |
| $C_{2v}$ | $\begin{pmatrix} 0 & g\kappa_I + \Gamma \cos\beta & \Delta \sin\beta \sin\alpha \\ -g\kappa_I - \Gamma \cos\beta & 0 & E \sin\beta \cos\alpha \\ -\Delta \sin\beta \sin\alpha & -E \sin\beta \cos\alpha & 0 \end{pmatrix}$ |

The only non-dissipative terms (proportional to $g$) correspond to the photovoltaic/Drude-type effect presented in Eq. 9. Indeed, $\boldsymbol{E}^{int}$ is parallel to $z$ in the ferroelectric phases and thus yields the term $\sigma_m^{xy} = g$ that provides an electric current that changes sense when $\boldsymbol{E}^{int}$ reverses. All the other terms are dissipative and therefore can be neglected at low beam intensities. It is interesting to note that these antisymmetric dissipative contributions are the same in the ferroelectric tetragonal $C_{4v}$ phase and in the paraelectric tetragonal $D_{4h}$ phase, and have the same sense in domains with opposite spontaneous polarizations. That indicates the essentially photoelectric character of these contributions, which reinforces their smallness, since they have a relativistic origin [16]. Reversing the sense of $\boldsymbol{k}$ reverses the sign of the corresponding current contributions.

When the beam travels parallel to $z$, the $x-z$ and $y-z$ components of $\sigma_m$ vanish. Accordingly, no additional current parallel to $z$ is provoked by the applied magnetic field. Conversely, when the beam travels normally to $z$, the contributions to $\sigma_m^{xy}$ that are independent on the domain (proportional to $\Gamma$) vanish. Thus, applying a magnetic field parallel to $x$ yields a current parallel to $y$, wich follows the sense of the spontaneous polarization (and reciprocally for $y$).

Table XI shows the forms of the symmetric contributions to $\sigma_m$ in the three phases.

**TABLE XI.** Symmetric contributions to the photomagnetic conductivity tensor in the three phases.

| Crystal point group | $\sigma_m$ |
|---|---|
| $D_{4h}$ | $\sin\beta \begin{pmatrix} 0 & 0 & D\sin\alpha \\ 0 & 0 & -D\cos\alpha \\ D\sin\alpha & -D\cos\alpha & 0 \end{pmatrix}$ |
| $C_{4v}$ | $\sin\beta \begin{pmatrix} 0 & 0 & D\sin\alpha \\ 0 & 0 & -D\cos\alpha \\ D\sin\alpha & -D\cos\alpha & 0 \end{pmatrix}$ |
| $C_{2v}$ | $\begin{pmatrix} 0 & B\cos\beta & D\sin\beta\sin\alpha \\ B\cos\beta & 0 & -F\sin\beta\cos\alpha \\ D\sin\beta\sin\alpha & -F\sin\beta\cos\alpha & 0 \end{pmatrix}$ |

All the contributions are dissipative and independent on the ferroelectric domains, betraying their photoelectric character and their relative smallness. The absence of diagonal terms shows that applying the field along any of the principal axes $x, y$ or $z$ yields a current normal to it. On the other hand, applying the wave with $\boldsymbol{k}$ parallel to $z$ cancels out the tensor in the two tetragonal phases. In these phases, applying the magnetic field parallel to $z$ produces a current parallel to $z$, whereas when the field is normal to $z$ and $\boldsymbol{k}$ in the $x-y$ plane yields a current parallel to $\boldsymbol{k}$.

Let us finally describe the photo-induced Hall effect in a rectangular barrel of the $C_{4v}$ phase. This shape permits to preserve the tetragonal symmetry and to apply our analysis at the macroscopic level.

When the wave propagates parallel to $z$ (Fig. 9a) $\boldsymbol{j}_{ph}$ is also parallel to $z$ (see Table IV) and the photomagnetic conductivity reads:

$$\sigma_m = \begin{pmatrix} 0 & \varsigma\kappa_I + \Gamma & 0 \\ -\varsigma\kappa_I - \Gamma & 0 & 0 \\ 0 & 0 & 0 \end{pmatrix} \quad (10)$$

Thus, applying the magnetic field parallel to $y$ yields an additional current contribution parallel to $x$. Since the barrel has a finite transversal size, its $y-z$ walls become electrically charged and thus create a transversal electric field $\boldsymbol{E}_{comp}$ that compensates the $x$ current. Apart from the small relativistic $\Gamma$ term (note that Eq. 10 holds for a wave travelling along $+z$, for the opposite beam sense $\Gamma$ should be replaced with $-\Gamma$) and the absence of applied DC voltage, this effect is exactly similar to the conventional Hall effect.

When the wave propagates parallel to $x$, $\boldsymbol{j}_{ph}$ is parallel to $y$ (see Table IV) and the photomagnetic conductivity reads:

$$\sigma_m = \begin{pmatrix} 0 & \varsigma\kappa_I & 0 \\ -\varsigma\kappa_I & 0 & -\Delta + D \\ 0 & \Delta - D & 0 \end{pmatrix} \quad (11)$$

Applying the magnetic field parallelly to $x$ (Fig. 9b) yields a current parallel to $y$ and the conventional Hall effect: the $x-z$ surface charge creates an internal electric field $\boldsymbol{E}_{comp}$ parallel to $y$ compensating both the photovoltaic and Hall currents. Applying $\boldsymbol{B}$ parallelly to $y$ (Fig. 9c) yields a current with a conventional Hall type contribution parallel to $x$ and a non-conventional contribution parallel to $z$ that does not need to be compensated.

At last, when the beam propagates in an arbitrary direction and the field is applied parallel to $x$ (or $y$), one finds conventional contributions parallel to the field $(x)$, and unconventional contribution normal to it $(y)$ and accordingly, compensating internal electric fields (Fig. 9d) in the $x-y$ plane. The non-conventional effects are dissipative and relativistic, so that their amplitudes must be very small.

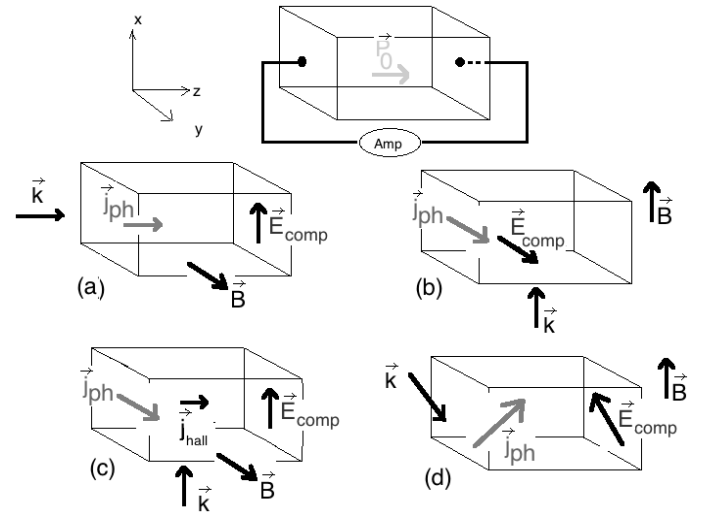

**FIG. 9**. Photo-Hall effect in a tetragonal barrel. The electric current can flow across the $x-y$ walls. The photo induced current in the $x-y$ plane is compensated by an internal electric field $E_{comp}$ due to the surface charges on the $y-z$ or $x-y$ walls. $j_{ph}$ is the photovoltaic current in the absence of an applied magnetic field. (a) wave vector beam $k$ parallel to $x$ and applied magnetic field $B$ parallel to y. (b) $k$ parallel to $x$ and $B$ parallel to $x$. (c) $k$ parallel to $x$ and $B$ parallel to $y$. (d) $k$ in an arbitrary direction and $B$ parallel to $x$.

## V. DISCUSSION AND CONCLUSION

We have calculated the responses to illumination corresponding to all types of vectors and second-order tensors in three phases with tetragonal/paraelectric, tetragonal/ferroelectric, and orthorhombic/ferroelectric symmetries. This works covers a vast domain of applications because many physical effects are accounted for by its results and because a large variety of materials are involved in these classes. Let us notice that even though any material with the right symmetry is concerned and will develop automatically the calculated tensors, their actual measurability depends on the type of material. For instance, the photovoltaic effect essentially concerns semiconductors with an adapted band gap, the magnetic effects involve materials containing magnetic ions in their crystalline cell or doped with such ions, and finally we expect a sufficient amplitude for measuring the photo-Hall effect only in metals. The governing parameter that allows to determine which material is susceptible to exhibit large photo-effects is its ability to transfer the incident photo energy to the electronic system. Along this line, efficient photovoltaic crystals are of course the main candidates. Within this family, the famous tetragonal ferroelectric perovskites BaTiO$_3$ and PbTiO$_3$ are well suited, at least for non-magnetic effects.

The phase transitions between the three studied structures play an important role that we should briefly discuss. We can see in Tables I—VIII that each symmetry breakdown ($D_{4h} \to C_{4v} \to C_{2v}$) yields additional terms in the induced tensors, corresponding to successive onset or enrichment of the associated physical properties. Each new coefficient $^A K_L^{mp}$ has a well-defined tensor character (which depends on the vector or tensor nature of $\Sigma$, but also on the value of $L$) allowing to consider it as a secondary (non-symmetry-breaking) order parameter (at the center of the Brillouin zone) coupled to the primary order parameter describing the transition [39]. The type of coupling between these two order parameters controls the temperature dependence of the onsetting $^A K_L^{mp}$ below the transition. This coupling has not a universal feature since the macroscopic symmetry breakdown sequence $D_{4h} \to C_{4v} \to C_{2v}$ can be associated with a number of associated space group sequences.

The simplest case arises when the first transition ($D_{4h} \to C_{4v}$) is a primary ferroelectric transition (i.e., the macroscopic polarization is the primary order parameter) and the second transition ($C_{4v} \to C_{2v}$) is primary ferroelastic (Fig. 10a), which are realized in the GKN family. Since these order parameters lie at the center of the Brillouin zone, their coupling with the coefficients $^A K_L^{mp}$ can be obtained by simple Clebsh-Gordan combinatorial rules. This is no longer the case when the transition order parameter is spanned by waves lying in the volume of the Brillouin zone, so that the cell volume is changed at the transition. This phenomenon occurs for instance in PKN. Indeed, at the $D_{4h} \to C_{2v}$ transition the elementary cell is doubled with $a_{ort} = a_{tet} + b_{tet}$ and $b_{ort} = a_{tet} - b_{tet}$, so that the orthorhombic cell is rotated by 45° with respect to the tetragonal cell (Fig. 10b). Consequently, the comparison between the two phases in Tables I—VIII has to be careful: the $x, y$ coordinates of the orthorhombic cell actually match with the $x + y, x - y$ coordinates of the tetragonal phase.

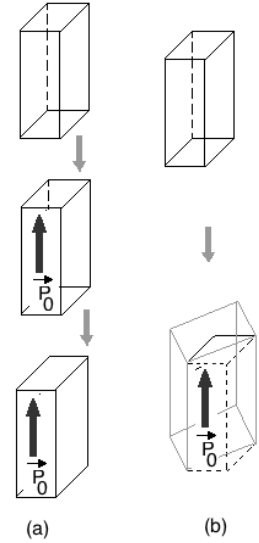

**FIG. 10.** Cell relationships between the three phases with symmetry groups D$_{4h}$, C$_{4v}$ and C$_{2v}$. (a) In the primary ferroelectric-ferroelastic model. (b) In PKN.

Let us now explore the role of the in-plane anisotropy by comparing some photo-effects in tetragonal ferroelectrics ($C_{4v}$) materials (PKN, GKN) with their analogues in trigonal ferroelectric ($C_{3v}$) materials (LiNbO$_3$, BiFeO$_3$). The comparison shows that the photovoltaic effect has precisely the same form to the first order in $L$ in the two classes of systems. However, we expect differences at higher orders since larger values of $L$ allow the crystal to better adapt its responses to the details of its anisotropy. Conversely, considering second-order tensors, such as the photomagnetoelectric tensor, shows strong differences between the tetragonal and trigonal systems.

In conclusion, we have systematically predicted the form of the angular variations of four vectorial and four tensorial responses of tetragonal and orthorhombic ferroelectric materials to linearly polarized light beam on using a new phenomenological method that cures some drawbacks of the traditional approach. We have applied our results to the phase sequence observed in the polymorphism of tetragonal tungsten bronzes. We discuss in detail some geometric aspects of the optical rectification, the photomagnetism, the photovoltaic effect, the photoelasticity, the photorefractive effect, the photomagnetoelectricity, the photoconductivity and the photo-Hall effect. We have paid special attention to two unconventional effects: (i) The photo-Hall effect, in which the magnetic field deviates photo-excited electrons instead of conducting electrons as in the conventional Hall effect, and (ii) a non-linear optic effect in which a primary light beam controls the rotation of the optic tensor that, in turn, refracts the extraordinary ray of a secondary beam. Both effects are precisely described by our approach with a small number of phenomenological parameters that can be easily determined on probing these effects.